\documentclass{elsart}
\usepackage{graphicx}
\usepackage{amsmath}
\usepackage{cite}

 \journal{  Journal of Colloid
and Interface Science \textbf{302} (2006) 605-612
\quad\quad\quad\quad\quad\quad\quad\quad\quad\quad\quad\quad\quad\quad\quad\quad\quad\quad\quad\quad\quad\quad\quad\quad}
\begin{document}
\begin{frontmatter}
\title{Modelling of the moving deformed triple contact line: influence of the fluid inertia}
\author{Vadim S. Nikolayev} \ead[email]{ vadim.nikolayev@espci.fr  http://www.pmmh.espci.fr/$\sim$vnikol}
 \address{ESEME, Service des Basses Temp\'eratures, CEA-Grenoble, France  and} \address{CEA-ESEME, ESPCI-PMMH,
10 rue Vauquelin, 75231 Paris Cedex 5, France}
\author{Sergey L. Gavrilyuk}
\address{University of Aix-Marseille \& Polytech Marseille UMR CNRS 6595 IUSTI,\\ 5 rue E. Fermi,
13453 Marseille Cedex 13, France
\ead[email]{sergey.gavrilyuk@polytech.univ-mrs.fr}}
\author{Henri Gouin}
\address{University of Aix-Marseille \& U.M.R.  C.N.R.S. 6181, \\ Case 322, Av. Escadrille
 Normandie-Niemen, 13397 Marseille Cedex 20 France\\
\ead{henri.gouin@univ-cezanne.fr}}
\begin{abstract}
For partial wetting, motion of the triple liquid-gas-solid contact line is influenced by heterogeneities of the
solid surface. This influence can be strong in the case of inertial (e.g. oscillation) flows where the line can
be pinned or move intermittently. A model that takes into account both surface defects and fluid inertia is
proposed. The viscous dissipation in the bulk of the fluid is assumed to be negligible as compared to the
dissipation in the vicinity of the contact line. The equations of motion and the boundary condition at the
contact line are derived from Hamilton's principle. The rapid capillary rise along a vertical inhomogeneous wall
is treated as an example.
\end{abstract}
\begin{keyword}
Wetting; spreading; pinning; wetting hysteresis
\end{keyword}
\end{frontmatter}

\section{Introduction}

In equilibrium, the wetting properties of a liquid in contact with a solid are well defined by the static
contact angle $\theta_{eq}$ \cite{deG}. In the case of complete wetting ($\theta_{eq}=0$), a thin prewetting
film exists in front of the bulk of the liquid and its dynamics is well understood too. For partial wetting
($\theta_{eq}\neq 0$), a difficulty arises when the triple solid-liquid-gas contact line is moving. Huh and
Scriven \cite{Huh} showed that the viscous flow in contact line vicinity cannot be described by the no-slip
condition, i.e. zero fluid velocity at the solid wall. Indeed, since the contact line belongs at the same time
to the liquid and to the solid, the contact line velocity is ambiguous. This leads to a nonphysical divergence
of the hydrodynamic pressure and of the viscous dissipation at the contact line. In reality, the dissipation is
anomalously large but finite.

A motion of the straight contact line is studied in the large majority of works. However, wetting dynamics
experiments at partial wetting are almost inevitably submitted to an influence of surface heterogeneities that
we call ``defects" for the sake of brevity. On small scale, they lead to the contact line deformation; on
macroscopic scale, to the contact angle hysteresis, i.e. to the difference between the advancing and receding
contact angles. The contact line elasticity approach was proposed to describe the static deformation of the
contact line by a localized defect \cite{deG}. It resulted in the logarithmic contact line shape, which was
shown to describe correctly experimental data in the intermediate range of distances from the defect center.
Obviously, the logarithmic shape fails to describe the contact line both very close to the defect and far from
it where the contact line deformation needs to be finite. To obtain the correct description in the whole range,
the influence of gravity \cite{EuLet03} or the fluid mass conservation \cite{PRE05} needs to be taken into
account.

While the static hysteresis is understood relatively well \cite{deG,Garoff,Gouin3}, the contact line dynamics in
the presence of defects is under active discussion, see \cite{Moulinet} for related references. The contact line
speed is often presented as a function of the dynamic contact angle $\theta$ and of the static contact angle.
However, an ambiguity appears in definition of this static value which should lie somewhere between the
advancing and receding values of the static contact angle.

Among theoretical studies of the dynamics in the presence of defects (that we call dynamics of the ``deformed
contact line") one can distinguish the papers \cite{JR,Raph1}. Joanny \& Robbins \cite{JR} proposed a general
formalism for an arbitrary contact line shape but solved it only for the straight contact line. Golestanian and
Rapha\"el \cite{Raph1} used the quasi-static dynamics in conjunction with the contact line elasticity model to
study the random deformation of the contact line.

To describe the effect of defects, a single phenomenological parameter $\xi$, the dissipation coefficient, was
introduced \cite{PRE02,EuLet03} following \cite{RCO99}. $\xi$ contains all the physics of the contact line
motion mechanism and has to be specific to the fluid-substrate system. To introduce it into the theory, one
needs to assume that (i) the viscous dissipation in the bulk of the fluid is neglected with respect to that in
the contact line vicinity, and (ii) the contact line dissipation is independent of the direction of the contact
line motion (advancing or receding). The energy dissipated per unit time at the contact line is then given in
the lowest order of the contact line velocity $v_n$ (measured in the direction normal to the contact line) by
the expression
\begin{equation}\label{diss}
T=\int\frac{\xi \, v_n^2}{2}\;{\rm d}l,
\end{equation}
where the integration is performed along the contact line. The defects are modelled by the spatial variation of
the surface energy of the support which, according to the Young-Laplace formula, is equivalent to the spatial
variation of $\theta_{eq}$. It was shown in \cite{PRE05} that in quasistatic approximation (i.e. when the fluid
surface is assumed to be at equilibrium shape at any time moment) the expression (\ref{diss}) leads to the
equation valid for any contact line (or fluid surface) geometry,
\begin{equation}
v_n = \frac{\sigma}{\xi}(\cos\theta_{eq}-\cos\theta).  \label{cos}
\end{equation}
where $\sigma$ is the surface tension. Therefore, in any contact line motion model that results in (\ref{cos}),
the coefficient $\xi$ can be interpreted as the contact line dissipation per its unit length. For instance,
$\xi$ can be directly assimilated to the friction coefficient of Blake and Haynes \cite{BH1}. The dissipation
coefficient theory allows the contact line dynamics and the dynamic advancing and receding angles to be
predicted once the surface heterogeneity and the initial conditions are defined \cite{CondMat05}.

The relationship (\ref{cos}) has been verified against multiple experiments (see \cite{Blake} for a review).
Using (\ref{cos}), the $\xi$ value can be obtained from them. According to various experimental data, $\xi$ is
much larger than the shear viscosity $\eta$, which corroborates the assumption (i) above. However, there is
still no certainty even about the order of value of $\xi/\eta$ for a particular liquid-solid system. The
experimental values range from 300 \cite{RCO99} to $10^7$ \cite{JFM02} and depend on the conditions of
experiment, the state of the substrate, etc.

It is argued recently \cite{Moulinet} that $v_n$ is non-linear in $\cos\theta$ which is equivalent to the
statement that $\xi$ varies strongly with $v_n$. At large $v_n$, the observed deviations from the behavior
(\ref{cos}) can be explained by the model \cite{Blake}. However, one needs to treat the experimental $\xi$ value
at small contact speeds with caution because it is usually determined \cite{Harmonic,Moulinet,Hamraoui,JFM02}
from the slope of the dependence of $\langle v_n\rangle$ on $\langle\cos\theta\rangle$ (where the angle brackets
mean averaging along the contact line). It is shown in \cite{CondMat05} that due to the collective effect of the
defects, the dependence (\ref{cos}) with \emph{constant} $\xi$ can result in a highly non-linear dependence of
$\langle v_n\rangle$ on $\langle\cos\theta\rangle$ near the pinning threshold where $\langle v_n\rangle$ is
small. The inferred from the slope $\xi$ value turns out to be larger than its actual value. The influence of
defects can thus lead to averaged $v_n(\theta)$ equations different from (\ref{cos}) at small contact line
speeds. To separate the effect of defects from the physics mechanism of the contact line motion, a description
of the dynamics of the contact line deformed by the defects is necessary.

Most theoretical approaches aim to describe the slow fluid motion and thus neglect the fluid inertia. However,
the inertial effects can be important in many common situations. One can mention the early stages of drop
spreading \cite{QuereCl2}, possibly after its impact with a support \cite{Roisman}, the capillary rise
\cite{QuereCl,Dreyer,Dreyer2}, and oscillating flows in various geometries: those of sessile or pendant drops
\cite{Lang04,Noblin,surfract,pendant,Lyubimov}, liquid in containers \cite{Hocking,Ting,Jiang,Nicolas}, or
capillary bridges \cite{bridges}.

In the partial wetting regime all these situations involve a period of time where $v_n$ is small and then the
defect impact on dynamics is large. Indeed, $v_n$ necessarily passes through zero during oscillations. As for
relaxation processes, the influence of defects becomes important at the end of the evolution.

This problem is especially important for oscillation flows where the choice of the boundary conditions at the
solid walls is a longstanding problem \cite{Hocking}. Evidently, when the amplitude of oscillations is small
enough, the contact line is pinned \cite{Noblin}, so that the fixed position boundary condition need to be
applied. When the amplitude is larger, the contact line can be pinned during a part of the oscillation period
\cite{Ting,Jiang}. At large amplitude, it moves almost all the time. In the relatively well studied
\cite{pendant,Nicolas} first regime, there is no singularity at the contact line. The transition from the first
to the second regime and especially the complicated dynamics in the second regime did not yet find its
explanation. In this paper we develop a model suitable to solve these problems.

The equations of motion are derived in the ``shallow water" approximation in sections \ref{secII} and
\ref{secIII}. This approximation permits to solve analytically a model problem (section \ref{secIV}) in order to
validate the model and find a first-order ``inertial" correction to the quasi-static behavior. The results are
summarized in the section \ref{secV}.

\section{Equations of motion}\label{secII}

The oscillating flows are considered often in the inviscid fluid approximation \cite{Nicolas,Lyubimov}. This is
justified by the thinness of the viscous boundary model at moderate Reynolds numbers where the flow is not yet
turbulent. For the contact line problems, an additional argument related to the anomalously strong contact line
dissipation can be put forward. Indeed the experiments \cite{Hocking,Ting,Noblin,Jiang} show that its
contribution can be larger than that of the viscosity in the liquid bulk. Therefore, we will neglect the viscous
dissipation in the bulk by assuming that the liquid is inviscid. The only source of dissipation is assumed to be
at the contact line.

\begin{figure}
  \begin{center}
  \includegraphics[height=4cm]{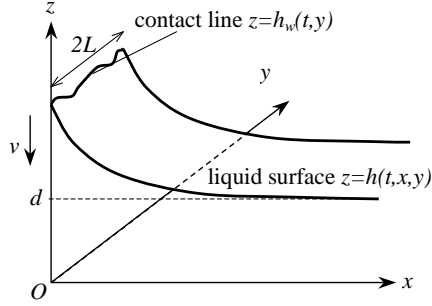}
  \end{center}
\caption{Sketch of the liquid layer near the vertical (possibly moving with the speed $v$) wall
with surface defects.} \label{Wilh}
\end{figure}

A Cartesian reference frame $Oxyz$ is used to describe the fluid motion in the domain $x>0$, $-\infty<y<\infty$,
$0<z<h(t,x,y)$ (Fig.~\ref{Wilh}). The free surface is described by the equation $z=h(t,x,y)$ and is assumed to
be weakly deformed
\begin{equation}
|\nabla h| \ll 1, \label{assumption}
\end{equation}
where the gradient is taken with respect to the variables $(x,y)$. The vertical wall ($x=0$) may move in $z$
direction. The gravity is directed downward. The contact line is described by the equation
\begin{equation}
z=h_{w}(t,y)=h(t,0,y).  \label{boundary}
\end{equation}

For convenience, the function $h(t,x,y)$ is assumed to be periodic (with the period $2L$) in the
$y$-direction. This is not a restrictive assumption since one might put $L\rightarrow\infty$ in the
resulting equations.

To derive the governing equations and boundary conditions in the absence of the volume dissipation, we use the
Hamilton principle of stationary action. The idea is to formulate the Lagrangian, and to present the variation
of the Hamilton action as a linear functional of virtual displacements allowing to find not only the governing
equations but also the natural boundary conditions \cite{Sedov,Berd}.

In the present work, we will use the ``shallow water" approximation which will allow the analytical results to
be obtained and analyzed. The Hamilton action $a$ and the Lagrangian $\mathcal{L}$ (which is the difference
between the kinetic and the potential energy of the system) read
\begin{eqnarray}
a=\int_{t_1}^{t_2}\mathcal{L}\,\textrm{d}t,\quad \mathcal{L}=\frac{1}{2L} \int_{0}^\infty
\textrm{d}x\int_{-L}^{L}\left( \frac{\rho h\left| \vec{u}\right| ^{2}}{2}-\frac{\rho gh^2}{2}\right)
\textrm{d}y- \nonumber\\ \frac{\sigma }{2L}\int_{0}^{\infty }\textrm{d}x\int_{-L}^{L}\left( \sqrt{1+|\nabla h|
^{2}}-1\right) \textrm{d}y+\frac{\sigma }{2L}\int_{-L}^{L}\textrm{d}y
\int_{0}^{h_{w}(t,y)}c(y,z)\,\textrm{d}z.\label{Lagrangian}
\end{eqnarray}
Here $\rho$ is the density of the fluid assumed incompressible, $\vec{u}=(u_x,u_y)$ is the fluid velocity field.
The first integral in the Lagrangian is a difference between the kinetic and the gravitational potential energy
of the system written in the shallow water approximation. The second integral is the liquid-gas interface
energy, and the last integral is the energy of the fluid-solid surface \cite{Garoff}, $\sigma c(y,z)$ being the
difference between the interfacial tension of gas-solid and liquid-solid interfaces. According to the Young
expression, $c(y,z)={\rm cos}(\theta_{eq})$, where $\theta_{eq}$ is the value of the equilibrium contact angle.
The variation of the latter along the substrate (i. e. along the $y-z$ plane) reflects the surface heterogeneity
(surface defects).

The variation of the action is taken under the mass conservation constraint
\begin{equation}
\frac{\partial h}{\partial t}+\nabla\cdot (h\vec{u})=0 \label{mass_conservation}
\end{equation}
and the non-penetration boundary condition at the vertical wall
\begin{equation}\label{cond0}
\left. u_x\right| _{x=0}=0.
\end{equation}

Search of the extremum for the action $a$ (see Appendix \ref{AA}) results in the following fluid motion
equation:
\begin{equation}
\frac{\partial \left( \rho h\vec{u}\right) }{\partial t}+\nabla\cdot (\rho h\vec{u}\otimes \vec{u})+ \rho
gh\nabla h-\sigma h\nabla(\Delta h)=0. \label{governing}
\end{equation}
Its back substitution into the Lagrangian (\ref{Lagrangian}) reduces the latter to the form (see also
(\ref{variation}))
\begin{equation}\label{Ltrunc}
 \delta\mathcal{L}=\frac{\sigma}{2L}
\int_{-L}^{L}\left[\left.\frac{\partial h}{\partial x}\right|_{x=0}+c(y,h_{w})\right]\delta
h_{w}\,\textrm{d}y.
\end{equation}

The dissipation is introduced \cite{EuLet03} in the usual way through the dissipation functional, (\ref{diss}).
For this, we need to find an expression for the normal velocity $v_n$ of the contact line. When the vertical
wall moves vertically with respect to the reference system with the velocity $v$, the velocity of the contact
line with respect to the wall reads
\begin{equation*}
v_{n}=\left(\frac{\partial h_{w}}{\partial t}+v\right)\left[1+\left( \frac{\partial h_{w}}{\partial y}\right)
^{2}\right]^{-1/2}\simeq \frac{\partial h_{w}}{\partial t}+v,
\end{equation*}
where $v$ is assumed to be positive for the downward plate motion. Since the volume dissipation inside the fluid
is neglected, the dissipation functional does not depend on the bulk fluid velocity $\vec{u}$ and the
dissipation account does not influence the governing equation (\ref{governing}).

Following \cite{EuLet03}, one can write now the generalized Euler-Lagrange equation which takes into account the
dissipation term
\begin{equation}
\frac{d}{dt}\left(\frac{\delta {\mathcal L}}{\delta\dot{h}_w}\right) -\frac{\delta {\mathcal L}}{\delta
h_w}=-\frac{\delta T}{\delta\dot{h}_w}, \label{Lagr}
\end{equation}
where $\delta\ldots/\delta\ldots$ means functional derivative and $\dot{h}_w$ denotes the time derivative. The
substitution of (\ref{Ltrunc}) into (\ref{Lagr}) results in the equation
\begin{equation}
\sigma \left[ \left.\frac{\partial h}{\partial x}\right|_{x=0}+c(y,h_{w})\right] = \xi
\left(\frac{\partial h_{w}}{\partial t}+v\right) \label{modif}
\end{equation}
that defines the contact line position $h_w$. Eq.~(\ref{modif}) plays a role of the boundary condition for the
momentum equation (\ref{governing}).

Let us note that the shallow water approximation is not necessary to obtain the main result (\ref{modif}). It is
known \cite{Berd} that if the full Lagrangian (involving the integration over $z$ of the kinetic energy) is used
instead of its ``shallow water" counterpart (\ref{Lagrangian}), the classical Euler equation of motion for the
inviscid fluid result is obtained instead of the equation (\ref{governing}). The back substitution into the
Lagrangian would lead to the same expression (\ref{Ltrunc}), which is necessary to obtain Eq.~(\ref{modif}).

\section{Reduction of governing equations}\label{secIII}

The boundary conditions for the governing equations (\ref{mass_conservation}, \ref{governing}) should be
completed by the condition at infinity
\begin{equation}\label{cond}
h|_{x\rightarrow\infty}=d,
\end{equation}
where $d$ is the constant depth. The following initial conditions will be used,
\begin{equation}\label{init}
\vec{u}|_{t=0}=0,\quad h|_{t=0}=d.
\end{equation}
The governing equations written for the deviation from the initial state $\hat{h}=h-d$ can be linearized:
\begin{equation}
\frac{\partial \hat{h}}{\partial t}+d\nabla\cdot \vec{u}=0,\quad \rho \frac{\partial \vec{u}}{\partial t}+\rho
g\nabla \hat{h}-\sigma \nabla\Delta \hat{h}=0. \label{linearised}
\end{equation}
The velocity $\vec{u}$ is eliminated from these equations by applying the time derivative to the first of them
and subtracting the divergence of the second. The result reads
\begin{subequations}\label{set}
\begin{eqnarray}
\rho\frac{\partial^2\hat{h}}{\partial t^2}-\rho gd\Delta \hat{h}+\sigma d\Delta^2\hat{h}=0.\label{master}
\end{eqnarray}
Evidently, the first time-dependent term of this equation is inertial. Two other terms correspond to the gravity
and to the surface tension contributions respectively. Without the first term, Eq.~(\ref{master}) would reduce
to the quasistatic linearized equation for the interface shape \cite{EuLet03}.

The boundary and initial conditions for (\ref{master}) are obtained from (\ref{cond0}, \ref{cond}--\ref{init})
by using (\ref{linearised}):
\begin{eqnarray}
\hat{h}|_{x\rightarrow\infty}=0, \label{cond_minf}\\
\left.\frac{\partial}{\partial x}\left(\rho g\hat{h}-\sigma \Delta\hat{h}\right)\right|_{x=0}=0,\label{condu0}\\
\hat{h}|_{t=0}=0, \label{condh0}\\
\left.\frac{\partial\hat{h}}{\partial t}\right|_{t=0}=0,\label{conddh0}
\end{eqnarray}
The nonlinear equation (\ref{modif}) rewritten for $\hat{h}$
\begin{eqnarray}
\frac{\partial \hat{h}_{w}}{\partial t}+v=\frac{\sigma}{\xi} \left[\left.\frac{\partial \hat{h}}{\partial
x}\right|_{x=0}+c(y,h_{w})\right], \label{hw}
\end{eqnarray}
\end{subequations}
where $\hat{h}_w=h_w-d$, closes the problem statement.

Only the solution for $h_w(t,y)$ is of interest. However, it can only be obtained by analyzing the full problem
(\ref{set}) for $h$. The set of equations (\ref{set}) allows $h_w$ to be obtained for arbitrary distribution of
the surface defects $c(y,z)$.

\section{Solution of a model problem}\label{secIV}

To demonstrate the applicability of the developed formalism, let us solve a problem of the
capillary rise on an immobile ($v=0$) wall with a single stripe defect (see Fig.~\ref{figh1} below)
\begin{equation}\label{cex}
  c(y)=\left\{\begin{array}{cc}
    c_d, & |y|\le w, \\
    c_s, & |y|> w,
  \end{array}\right.
\end{equation}
where $c_d,c_s$ and $w$ are constants, i.e. for the simplest case in which $c$ is independent of $z$. The values
of $c_d$ and $c_s$ are the cosines of the equilibrium contact angles inside and outside the stripe defect
respectively, $w$ being its half-width. This problem has already been solved in quasi-static approximation
\cite{EuLet03} (in which the fluid motion is neglected). We choose the same setup to show the role of the fluid
inertia by comparing the quasi-static and inertial solutions.

Since $z$ dependence in (\ref{cex}) is absent, the problem (\ref{set}) becomes linear. Following \cite{EuLet03},
$h$ and $h_w$ are broken into parts:
\begin{equation}\label{sum}\begin{array}{c}
  h=d+h_0(t,x)+h_1(t,x,y), \\
  h_w=d+h_{w0}(t)+h_{w1}(t,y), \\
\end{array}
\end{equation}
where $h_0$ and $h_{w0}$ are the averaged over the $y$-direction values of $\hat{h}$ and
$\hat{h}_{w}$ respectively. The problems for $h_0$ and $h_1$ are decoupled and can be solved
separately.

\subsection{Solution for $h_{w0}$}\label{subs-h0}

Consider first the problem for $h_0$ that can be obtained from
(\ref{set}) by accounting for the fact $\partial h_0/\partial
y\equiv 0$. By applying the Laplace transform
\begin{equation}\label{Lh}
\bar{h}_0(x)=\int_0^\infty h_0(t,x)\exp(-st)\,\textrm{d}t,
\end{equation}
one obtains from (\ref{set}\textit{a,d,e}) the ordinary
differential equation
\begin{equation}\label{eh0} \frac{d^4\bar{h}_0}{d
x^4}-2\alpha\frac{d^2\bar{h}_0}{d x^2}+\beta\bar{h}_0=0,
\end{equation}
where $\alpha=\rho g/(2\sigma)$, $\beta=\rho s^2/(\sigma d)$. The general solution of (\ref{eh0})
has a form\begin{equation}\label{Lh0}
\bar{h}_0(x)=a_1\exp(-p_1x)+a_2\exp(-p_2x)+a_3\exp(p_1x)+a_4\exp(p_2x),
\end{equation}
where
\begin{equation}\label{roots}
p_{1,2}=\sqrt{\alpha\pm\sqrt{\alpha^2-\beta}},\quad \mbox{Re}\;p_{1,2}\ge 0.
\end{equation}
The boundary conditions for $\bar{h}_0$ have the same form (\ref{set}b,c) as for $h_0$; (\ref{hw}) transforms
into
\begin{equation}\label{hw0}
s\bar{h}_{w0}=\frac{\sigma}{\xi} \left( \left.\frac{d \bar{h}_0}{d
x}\right|_{x=0}+\frac{c_s}{s}\right),
\end{equation}
where $\bar{h}_{w0}$ is the Laplace transform of $h_{w0}$. Because
of (\ref{cond_minf}),
\begin{equation}\label{a34} a_3=a_4=0,
\end{equation}
and (\ref{condu0}) results in the following condition
\begin{equation}\label{dhw0}
\left.\frac{d \bar{h}_0}{d x}\right|_{x=0}=-\bar{h}_{w0}(p_1+p_2)=
-\bar{h}_{w0}\sqrt{2\left(\alpha+\sqrt{\beta}\right)}.
\end{equation}
The final form for $\bar{h}_{w0}$ can be obtained by the substitution of (\ref{dhw0}) into (\ref{hw0}):
\begin{equation}\label{hw0L}
  \bar{h}_{w0}=c_s\frac{\sigma}{\xi}\left[s\left(s+\tau_r^{-1}
  \sqrt{s\sqrt{2\tau_i}+1}\right)\right]^{-1},
\end{equation}
where  $\tau_r=l_c\xi/\sigma$ is a characteristic time scale for the quasi-static relaxation \cite{EuLet03} and
$\tau_i=l_c/\sqrt{g d}$ is the characteristic inertial time, $l_c=\sqrt{\sigma/(\rho g)}$ being the capillary
length. The inverse Laplace transform of (\ref{hw0L}) can be found rigorously as described in Appendix. It reads
\begin{eqnarray}\label{hw0f}
h_{w0}(t)=c_sl_c\bigg\{1-\mbox{erfc}\left(\sqrt{\frac{t}{2\tau_i}}\right)+\nonumber\\
\frac{1}{2\sqrt{1+r^2}} \bigg\{\exp\left[\frac{t}{\tau_r}\left(r+\sqrt{1+r^2}\right)\right]
\mbox{erfc}\left[\sqrt{\frac{t}{2\tau_i}}\left(r+\sqrt{1+r^2}\right)\right]-\nonumber\\\exp\bigg[\frac{t}{\tau_r}\bigg(r-\sqrt{1+r^2}\bigg)\bigg]
\mbox{erfc}\left[\sqrt{\frac{t}{2\tau_i}}\left(r-\sqrt{1+r^2}\right)\right]\bigg\}\bigg\},
\end{eqnarray}
where $\mbox{erfc}(\cdot)$ is the complementary error function \cite{Bateman}, and $r=\tau_i/\tau_r$
characterizes a relative importance of the fluid inertia. It is related to the Weber number which can be
obtained from (\ref{master}) where both the terms of inertia and of surface tension are present. Taking $\tau_r$
and $l_c$ as the scales of time and length, one can obtain for these terms $\rho h/\tau_r^2$ and $\sigma
dh/l_c^4$. Their ratio is the Weber number, $We=r^2$.

\begin{figure}
  \begin{center}
  \includegraphics[height=6cm]{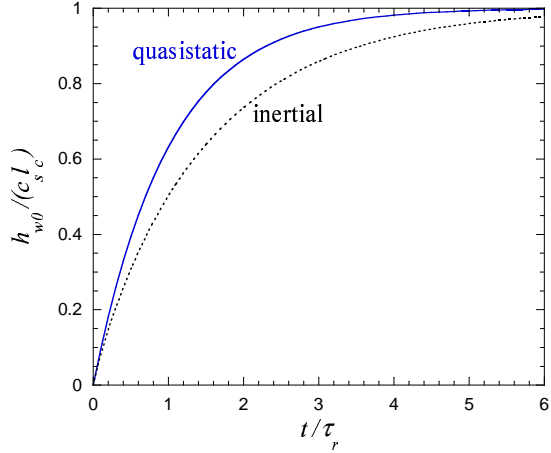}
  \end{center}
\caption{ The time dependence of the $y$-averaged hight of the contact line rise $h_{w0}$.  The exponential
quasi-static law (\ref{h0r}) (solid line) is compared to the inertial behavior (\ref{hw0f}) (dotted line)
calculated for $r=0.5$.} \label{figh0}
\end{figure}

In the limit $r\rightarrow 0$, (\ref{hw0f}) reduces to the quasi-static result \cite{EuLet03}
\begin{equation}\label{h0r}
h_{w0}(t)|_{r\rightarrow 0}=c_sl_c[1-\exp(-{t}/{\tau_r})].
\end{equation}
The difference between these two dependencies is transparent from Fig.~\ref{figh0}: the fluid inertia slightly
slows down the contact line relaxation. For $r\le 0.01$, the difference between the functions (\ref{hw0f}) and
(\ref{h0r}) is very small so that in practice the exponential solution (\ref{h0r}) is valid.

\subsection{Solution for $h_{w1}$}

The statement of the problem for $h_{1}$ is given by (\ref{set})
with an additional condition of zero $y$-averaged value of
$h_{1}$. For a single defect, this condition reduces to
$h_{1}(y\rightarrow\pm\infty)=0$, which allows the Fourier
transform
\begin{equation}\label{Fh} \tilde{h}_1(t,x)=\int_{-\infty}^\infty
h_1(t,x,y)\exp(-iky)\,\textrm{d}y
\end{equation}
to be applied. The Laplace transform (\ref{Lh}) can further be applied to $\tilde{h}_1(t,x)$ and
results in $\bar{h}_1(x)$.

\begin{figure}
  \begin{center}
  \includegraphics[height=6cm]{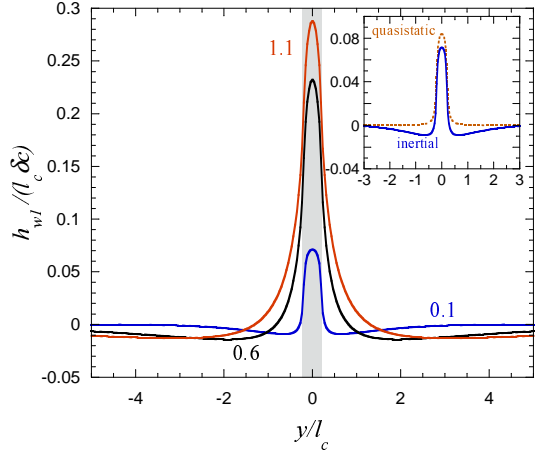}
  \end{center}
\caption{The contact line distortion $h_{w1}$ calculated for $r=0.5$. The defect area (half-width of which is
$w=0.2l_c$) with $\cos\theta_{eq}=c_d$  is shadowed. The rest of the plane has $\cos\theta_{eq}=c_s$. The
parameter of the curves is the time $t$ in the $\tau_r$ units. In the insert, the result of the inertial theory
for $r=0.5$ (solid curve) is compared to the quasi-static ($r\rightarrow 0$) result (dotted curve). Both curves
are calculated for $t=0.1\tau_r$.}\label{figh1}
\end{figure}

The further procedure is fully analogous to that developed in the previous section. It leads to the
following equation for $\bar{h}_{w1}$
\begin{equation}\label{hw1}
s\bar{h}_{w1}=\frac{\sigma}{\xi} \left( \left.\frac{d \bar{h}_1}{d
x}\right|_{x=0}+\frac{\tilde{c}_1}{s}\right),
\end{equation}
where $\tilde{c}_1$ is the Fourier transform of
$c(y)-c_s$,\begin{equation}\label{ct}
 \tilde{c}_1=2\delta c\frac{\sin kw}{k},
\end{equation}
and $\delta c=c_d-c_s$. The $\bar{h}_{1}$ derivative is related to
$\bar{h}_{w1}$ through
\begin{equation}\label{dhw1}
\left.\frac{d \bar{h}_1}{d
x}\right|_{x=0}=-\bar{h}_{w1}\frac{p_1p_2(p_1+p_2)}{k^2+p_1p_2},
\end{equation}
where $p_{1,2}$ are defined by (\ref{roots}), in
which
\begin{equation}\label{albe}
    \alpha=\frac{\rho g}{2\sigma}+k^2,\quad\beta=\frac{\rho }{\sigma d}s^2+\frac{\rho
    g}{\sigma}k^2+k^4
\end{equation}
should now be used. The substitution of (\ref{dhw1}) into (\ref{hw1}) leads to an expression, for which the
inverse Laplace transform is difficult to find. To overcome this difficulty, one can make use of the smallness
of the parameter $r$, i.e. assume the smallness of the contribution of the inertial effects. In this limit,
(\ref{dhw1}) reduces to
\begin{equation}\label{dhw1s}
\left.\frac{d \bar{h}_1}{d
x}\right|_{x=0}=-\bar{h}_{w1}\left[\sqrt{k^2+l_c^{-2}}+B(r\tau_rs)^2\right],
\end{equation}
where the coefficient $B$ is\begin{equation}\label{B}
    B=\frac{k^2+l_c^{-2}}{|k|}-\frac{k^2+1.5l_c^{-2}}{\sqrt{k^2+l_c^{-2}}}.
\end{equation}
Note that $B$ is positive and diverges at small $k$ like $B\sim
|k|^{-1}$. The equation for $\bar{h}_{w1}$ becomes
\begin{equation}\label{hw1L}
  \bar{h}_{w1}=\tilde{c}_1\left[s\left(\sqrt{k^2+l_c^{-2}}+l_c^{-1}\tau_rs+B(r\tau_rs)^2\right)\right]^{-1}.
\end{equation}
This expression is rational in $s$ and its inverse Laplace transform can be found using
conventional methods:
\begin{equation}\label{hw1f}
  \tilde{h}_{w1}(t)=\tilde{c}_1\left\{\frac{1}{\sqrt{k^2+l_c^{-2}}}+\frac{1}{B(r\tau_r)^2(s_1-s_2)}
  \left[\frac{\exp(s_1t)}{s_1}-\frac{\exp(s_2t)}{s_2}\right]\right\}.
\end{equation}
Here \begin{equation}\label{s12}
s_{1,2}=\frac{1\pm\sqrt{1-4Bl_cr^2\sqrt{k^2l_c^{2}+1}}}{2Bl_cr^2\tau_r}.
\end{equation}
To obtain the time-varying contact line shape $h_{w1}(t,y)$, one needs to apply the inverse Fourier transform to
(\ref{hw1f}). It can be done numerically. The result is shown in Fig.~\ref{figh1}.

At $r\rightarrow 0$, one recovers the quasi-static result \cite{EuLet03}
\begin{equation}\label{h1r}
\tilde{h}_{w1}(t)|_{r\rightarrow
0}=\frac{\tilde{c}_1}{\sqrt{k^2+l_c^{-2}}}\left[1-\exp\left(-\frac{t}{\tau_r}\sqrt{k^2l_c^2+1}\right)\right],
\end{equation}
which corresponds to a non-zero value at $k\rightarrow 0$. However, by considering the small $k$ limit of
(\ref{hw1f}), one arrives to the result \footnote{Notice that although $s_{1,2}$ become complex at small $k$,
$\tilde{h}_{w1}$ remains real.}
\begin{equation}\label{ll}
   \tilde{h}_{w1}(t)|_{k\rightarrow
0}=\frac{\tilde{c}_1t^2}{2B\tau_i^2}\rightarrow 0.
\end{equation}
This small $k$ ambiguity leads to a small but regular numerical error in $h_{w1}(t,y)$ when the inverse Fourier
transform of (\ref{hw1f}) is found at very small $r$. In practice, the expression (\ref{h1r}) should be used
instead for $r\le 0.05$.

Similarly to the case of ${h}_{w0}$, the inertial slowing down manifests itself in the case of ${h}_{w1}$. The
insert in Fig.~\ref{figh1} shows that the topmost point of the contact line rises slower when the hydrodynamic
motion is taken into account. However, the hydrodynamic effects are not limited to the simple slowing down, they
change the shape of the contact line deformation. In the beginning of the capillary rise, the joint effect of
the inertia and fluid mass conservation creates two wells in the ${h}_{1}$ shape (Fig.~\ref{figh1}) from which
the fluid flows off to form a bump in the middle. Later on, these cavities become more and more shallow and wide
until they disappear in the equilibrium static contact line shape \cite{EuLet03}.

\section{Results and discussion}\label{secV}

The main result of the present paper is a model that allowed to describe the inertial regimes of the contact
line motion whenever the impact of surface defects is important in the partial wetting regime, e.g. the
pinning-depinning of the contact line during its oscillation. The equations of motion were derived for the plane
vertical wall geometry and can be applied for the arbitrary defect pattern and both for the spontaneous and
forced contact line motion. The hydrodynamic shallow water approximation was used to obtain the solution in the
closed analytical form. However, this approach can be easily generalized to any geometry and fluid depth.

Indeed, let us have a closer look to the equation (\ref{modif}). Its most interesting feature is its invariance.
In the hydrodynamic theory, it has exactly the same form as in the quasi-static theory \cite{CondMat05}. Since
$c(y,z)=\cos[\theta_{eq}(y,z)]$ and $\partial h/\partial x|_{x=0}\simeq-\cos\theta$ under the assumption
(\ref{assumption}), one notices that (\ref{modif}) is nothing else than the equation (\ref{cos}). It was derived
under the assumption of the small deformation of the straight contact line. By analogy with the quasistatic
result \cite{PRE05}, it is however reasonable to assume that (\ref{cos}) is valid for any contact line
deformation. Any contact line inertial problem can thus be solved by applying the Euler set of equations for the
inviscid fluid with the boundary condition (\ref{cos}) where $\theta_{eq}(y,z)$ models the surface
heterogeneity.

The asymptotics of the capillary rise at small times which follows from the present model can be compared to the
available experimental data. It stems from the equation (\ref{cos}) that $v_n$ is constant at $t=0$ because
$\theta(t=0)$ is fixed by the initial condition for $h$. Since $h(t=0)=0$ has been chosen, $\theta(t=0)=\pi/2$
and $h(t)\sim t$ at small times independently of the geometry of the problem. This asymptotics agrees with the
results \cite{Dreyer} on the capillary rise between flat walls. The asymptotics $h(t)\sim t^{3/2}$ obtained for
the complete wetting case \cite{Quere,QuereCl} cannot be compared to our model restricted to the partial
wetting. As a matter of fact, due to existence of the prewetting film \cite{deG}, the contact line dissipation
anomaly is much weaker for the complete wetting. The kinetics of the capillary rise is thus defined by the
balance of the surface tension and inertia \cite{QuereCl} which results in $h(t)\sim t^{3/2}$.

The effect of inertia on the contact line relaxation is two-fold. First, it is the slowing down of the average
contact line motion. When the contact line speed is inhomogeneous, the inertia amplifies its variation along the
contact line.

The impact of the inertia on the contact line relaxation can be measured by the value $r=\tau_i/\tau_r$ of the
ratio of two characteristic times. The first of them is the inertial time and the second is the quasi-static
relaxation time. The quasi-static result is recovered in the limit $r\rightarrow 0$. Note that the Weber number
(the ratio of the inertial and surface tension terms) is related to $r$: $We=r^2$.

The simplicity of the developed approach is its main advantage. It uses a single phenomenological parameter (the
contact line dissipation coefficient $\xi$) and allows complicated three dimensional problems to be treated. In
this article we analyzed the contact line motion in presence of a stripe defect. This model can be directly
verified against experiments \cite{Nad,Caz,Limat}. However, the information presented in these papers is
insufficient for a direct comparison and additional measurements are necessary.

\begin{ack}
VN is indebted to D. Beysens for stimulating discussions.
\end{ack}

\appendix

\section{Derivation of the governing equation (\ref{governing})}\label{AA}

Assumption (\ref{assumption}) permits us to rewrite (\ref{Lagrangian}) in a simpler form
\begin{eqnarray}
\mathcal{L}=\frac{1}{2L}\int_{0}^\infty \textrm{d}x\int_{-L}^{L}\left( \frac{\rho h\left|
\vec{u}\right| ^{2}}{2}-\frac{\rho gh^{2}}{2}\right) \textrm{d}y-\nonumber\\
\frac{\sigma }{2L}\int_{0}^{\infty }\textrm{d}x\int_{-L}^{L}\frac{\left| \nabla h\right|
^{2}}{2}\,\textrm{d}y+\frac{\sigma }{2L}\int_{-L}^{L}\textrm{d}y
\int_{0}^{h_{w}(t,y)}c(y,z)\,\textrm{d}z\label{Lagrangian1}
\end{eqnarray}
The variation of the Lagrangian (\ref{Lagrangian1}) submitted to the constraint
(\ref{mass_conservation}) is taken in several steps. First, we introduce a smooth one-parameter
family of virtual motions
\begin{equation*}
\vec{x}=\vec{\Phi }(t,\vec{X},\varepsilon ),\quad \vec{x}=(x,y)
\end{equation*}
($\vec{X}$\ stands for the Lagrangian coordinates, $\varepsilon $ is a small parameter in the vicinity of zero
and $\vec{x}={\mbox{\boldmath $\varphi$}}(t,\vec{X})=\vec{\Phi }(t,\vec{X},0)$ is the real motion). Then, we
define the virtual displacements as functions of $(t,\vec{X})$:
\begin{equation*}
\delta \vec{x}(t,\vec{X})=\left. \frac{\partial\vec{\Phi}(t,\vec{X},\varepsilon )}{\partial \varepsilon }\right|
_{\varepsilon =0}.
\end{equation*}
Later, we  consider the virtual displacements as functions of ($t,\vec{x}$) due to the possibility to invert
$\vec{X}$ through $\vec{X}={\mbox{\boldmath $\varphi$}} ^{-1}(t,\vec{x})$. The variations of the free surface
$h(t,\vec{x})$ and the velocity field $\vec{u}(t,\vec{x})$ compatible with the mass conservation law
(\ref{mass_conservation}) are given in the motion space ($t,\vec{x}$) by (see, for example, \cite{Berd} or
\cite{Gav1})
\begin{equation}
\delta h(t,\vec{x})=-\nabla\cdot (h\delta\vec{x}),\quad \delta
\vec{u}(t,\vec{x})=\frac{d\delta\vec{x}}{dt}-\frac{\partial \vec{u}}{\partial \vec{x}}\delta \vec{x}.
\label{variations}
\end{equation}
Here
\begin{equation*}
\frac{d}{dt}=\frac{\partial }{\partial t}+\vec{u}\cdot\nabla
\end{equation*}
is the material time derivative. Finally, the expression of the variation of $a$ in terms of the
virtual displacements reads
\begin{eqnarray*}
2L\delta a=\int_{t_{1}}^{t_{2}}\textrm{d}t\int_{0}^{\infty }\textrm{d}x\int_{-L}^{L}\bigg[ - \nabla\cdot(h\delta
\vec{x})\frac{\rho \left| \vec{u}\right|^2}{2}+\rho h\vec{u}\cdot \left( \frac{d\delta
\vec{x}}{dt}-\frac{\partial
\vec{u}}{\partial \vec{x}}\delta \vec{x}\right) +\\
\rho gh\nabla\cdot (h\delta \vec{x})\bigg] \textrm{d}y+\sigma \int_{t_{1}}^{t_{2}}dt\int_{0}^{\infty
}\textrm{d}x\int_{-L}^{L} \nabla h\cdot \nabla\left[
\nabla\cdot (h\delta \vec{x})\right] \textrm{d}y+\\
\sigma\int_{t_{1}}^{t_{2}}\textrm{d}t\int_{-L}^{L}c(y,h_{w})\delta h_{w}\,\textrm{d}y.
\end{eqnarray*}
By integrating by parts, accounting for the periodicity in $y$-direction, and the boundary
condition (\ref{cond0}) in the form
\begin{equation*}
\left. \delta \vec{x}\cdot\vec{n}\right| _{x=0}=0,
\end{equation*}
where $\vec{n}=(1,0)$ is the unit normal vector to the plane $x=0$, we obtain
\begin{eqnarray*}
2L\delta a=\int_{t_{1}}^{t_{2}}\textrm{d}t\int_{0}^{\infty }\textrm{d}x\int_{-L}^{L} \left[-\frac{\partial
\left( \rho h\vec{u}\right)}{\partial t}- \nabla\cdot (\rho h\vec{u}\otimes\vec{u})-h\nabla(\rho
gh)+\sigma h \nabla(\Delta h)\right] \cdot \delta \vec{x}\textrm{d}y+\\
\sigma \int_{t_{1}}^{t_{2}}\textrm{d}t\int_{0}^{\infty }\textrm{d}x\int_{-L}^{L} \nabla\cdot\left[ \nabla\cdot
(h\delta \vec{x})\nabla h\right]\textrm{d}y+\sigma\int_{t_{1}}^{t_{2}}\textrm{d}t\int_{-L}^{L}c(t,y,h_{w})\delta
h_{w}\textrm{d}y,
\end{eqnarray*}
where $\otimes$ denotes the direct product of the vectors. The definition of $\delta h_{w}$
\begin{equation*}
\delta h_{w}=-\left. \nabla\cdot(h\delta \vec{x})\right|_{x=0}
\end{equation*}
gives us the final expression for the variation of the Hamilton action:
\begin{eqnarray}
2L\delta a=\int_{t_{1}}^{t_{2}}\textrm{d}t\int_{0}^{\infty }\textrm{d}x\int_{-L}^{L}\bigg[-\frac{\partial \left(
\rho h\vec{u}\right) }{\partial t}-\nabla \cdot (\rho h\vec{u}\otimes\vec{u})-h\nabla(\rho
gh)\nonumber\\
+\sigma h\nabla(\Delta h)\bigg]\cdot\delta\vec{x}\,\textrm{d}y+\sigma
\int_{t_{1}}^{t_{2}}\textrm{d}t\int_{-L}^{L}[\nabla h\cdot \vec{n}+c(y,h_{w})]\delta
h_{w}\,\textrm{d}y\label{variation}.
\end{eqnarray}
The variations $\delta\vec{x}$ and $\delta h_{w}$ are independent and the term containing $\delta\vec{x}$ is
absent in the dissipation function because of neglect of the viscosity in the bulk of the fluid. Therefore, the
integral containing $\delta\vec{x}$ in (\ref{variation}) has to be equal to zero, which results in the governing
equation (\ref{governing}).

\section{Laplace-inversion of (\ref{hw0L})}\label{appA}

The expression (\ref{hw0L}) has a form
\begin{equation}\label{af}
\bar{f}(s)=[s(s+2b\sqrt{s+m^2})]^{-1},
\end{equation}
for which the inverse Laplace transform is sought. The inverse of
\begin{equation}\label{afsh}
\bar{f}(s-m^2)=\frac{1}{2m^2b}\left[\frac{1/2}{\sqrt{s}-m}
+\frac{1/2}{\sqrt{s}+m}-\frac{u_2(u_2-u_1)^{-1}}{\sqrt{s}-u_1}-\frac{u_1(u_1-u_2)^{-1}}{\sqrt{s}-u_2}\right],\end{equation}
where
\begin{equation}\label{ar}
u_{1,2}=-b\pm\sqrt{b^2+m^2},
\end{equation}
is found using the inverse of $(\sqrt{s}+m)^{-1}$ listed in the tables \cite{Bateman}. To find the inverse of
$\bar{f}(s)$, the original of $\bar{f}(s-m^2)$ should be multiplied by $\exp(-m^2t)$ according to the rules of
the Laplace transform.

\bibliographystyle{elsart-num}

\bibliography{HydroWet_JCIS06}

\end{document}